\documentclass[epj,referee]{svjour}
\usepackage{graphicx}
\usepackage{amsmath}
\begin{document}
\title{Electroabsorption study of index-defined semiconducting carbon nanotubes}
\subtitle{A direct probe into carbon nanotube excitonic states}
\author{N. Izard\inst{1,2,}\thanks{\email{nicolas.izard@u-psud.fr}} \and
E. Gaufr\`es\inst{1}
\and X. Le Roux\inst{1} \and S. Kazaoui\inst{2}
\and Y. Murakami\inst{3} \and D. Marris-Morini\inst{1}
\and E. Cassan\inst{1} \and S. Maruyama\inst{4} \and L. Vivien\inst{1}
}
\institute{Institut d'Electronique Fondamentale, CNRS-UMR 8622, Univ. Paris-Sud,
91405 Orsay, France \and National Institute of Advance Industrial Science and
Technology (AIST), Tsukuba 305-8565 (Japan) \and Global Edge Institute, Tokyo
Institute of Technology, Tokyo, Japan \and Department of Mechanical Engineering,
The University of Tokyo, Tokyo, Japan}
\date{Received: date / Revised version: date}
\abstract{
Electroabsorption spectroscopy of well identified index-defined semiconducting
carbon nanotubes is reported. The measurement of high definition
electroabsorption spectrum allows direct indexation with unique nanotube
chirality. Results show that at least for a limited range of diameter, electroabsorption is directly proportional to
the exciton binding energy of nanotubes. Electroabsorption is a powerful
technique which directly probes into carbon nanotubes excitonic states, and may
become an useful tool for in-situ study of excitons in future nanotube based
photonic devices such as electroabsorption modulators.
\keywords{nanotube ; electroabsorption ; modulator}}
\maketitle
\section{Introduction}
\label{intro}
Single Wall Carbon Nanotubes (SWNT) are nano-sized unidimensional structures
which have been considerably studied in the past ten years for their very
peculiar electrical and optical properties. Semiconducting single-wall carbon
nanotubes (s-SWNT) with their direct band-gap, present a special interest for
nanophotonic and electro-optic applications\cite{PSSB-Avouris,apl-Gaufres}.
Electron confinement in one dimensional nanotube leads to strong electron-hole
localization with binding energy as high as 0.5~eV, and the formation of the
so-called excitonic states\cite{Science-Wang}. Exciton binding energy strongly
depends on the nanotube intrinsic parameters (diameter,
chirality)\cite{prb-Lebedkin}, and on external parameters such as
strain\cite{prl-Yang,nano-Leeuw} or dielectric constant of the surrounding
medium\cite{apa-Lefebvre,prb-Izard}. Therefore, the environnement can be used to
tailor s-SWNT optical properties; which could strongly affect s-SWNT properties,
making them very efficient sensors\cite{Science-Kong}. As excitonic processes
are fundamental, numerous studies buckle down to elucidate
them\cite{prb-Maultzsch,nano-Mohite,nano-Lefebvre}. Usually,
excitons could be classified under two categories, dipole-allowed ``bright''
excitons, and dipole-forbidden ``dark'' excitons, the former being probed by
linear absorption spectra while the latter are probed by two-photon absorption
spectroscopy\cite{Science-Wang}. In contrast, electroabsorption (EA)
spectroscopy should allow access to full nanotube excitonic states, and would be
directly related to excitonic binding energy\cite{prl-Zhao,nano-Perebeinos}.

Experimentally, EA spectroscopy of carbon nanotubes is quite challenging due to
the mixing of metallic and semiconducting nanotubes, and the mutually overlapped
absorptions of different chirality tubes having proximate excitonic
energies\cite{apl-Takenobu,nano-Gadermaier,prl-Kishida}. However, recent advances in SWNT separation
techniques, either by density gradient
ultracentrifugation\cite{nano-Arnold,naturenano-Ghosh} or by polymer-assisted
extraction\cite{Naturenano-Nish,nano-Chen}, allows to obtain well defined
semiconducting nanotubes with a limited extent of chiral index, without
detectable traces of metallic nanotubes\cite{apl-Izard}. In this paper, we
report the first electroabsorption experiments performed on well identified
index-defined semiconducting tubes, which give a direct insight into s-SWNT
electronic structure and excitonic states.

\section{Materials and Experimental details}
\label{exp}
The current study uses two different kinds of s-SWNT, prepared using a
polymer-assisted selective extraction technique\cite{apl-Izard}. The first
sample was made to be as close as possible to a single index nanotube
distribution. A narrow diameter SWNT distribution synthesized by the CoMoCAT
process (from SouthWest NanoTech. Inc.) was used as a starting material. This
raw material was dispersed with poly-9,9-di-n-octyl-fluorenyl-2,7-diyl (PFO)
polymer in a toluene / acetic acid solution using sonication which, after an
ultracentrifugation step, leads to sample ``CoMoCAT/PFO'' made of almost
exclusively of (7,5) chirality s-SWNT\cite{prb-Murakami}.

The second sample was made such as to have the widest possible range of s-SWNT
chiralities, but at the same time having a distribution of well-defined single
chirality peaks. For that purpose, a large diameter SWNT distribution of
as-prepared HiPCO SWNT powder (from Unidym Inc.) was used as a starting
material. Details of this sample preparation procedure are reported
elsewhere\cite{apl-Izard}. Briefly, the preparation procedure involves a
homogeneization step (1~h with a water-bath sonicator and 15~min with a tip
sonicator) of the PFO-SWNT mixture, followed by an ultracentrifugation step
(2~h, 150.000~g). This method leads to sample ``HiPCO/PFO'', with a majority of
(8,6), (8,7) and (7,6) s-SWNT chiralities, as well as few (7,5) and (9,7)
chiralities\cite{OptLett-Gaufres,OptExp-Gaufres}.

Both samples were drop casted on a 2x2~mm quartz substrate, with 100~$\mu$m
spaced interdigitated Cr/Au electrodes prealably deposited, to form thin layers.
Those s-SWNT-doped PFO thin layers were subsequently annealed at 180$^\circ$C
for 15~min, resulting in highly fluorescent layer, with homogeneous s-SWNT
density\cite{apl-Gaufres,SynthMet-Campoy}.

Electromodulation spectroscopy experiments were then
made\cite{prb-Aspnes,prl-Campbell}. The sample was set inside a vacuum chamber
($10^{-5}$ torr) to prevent air breakdown under high electric fields. The light
source was a high power Halogen/Xenon white lamp, with a Nikkon monochromator.
Electrical modulation was assured by a function generator at a frequency $f$ of
333~Hz, with a voltage amplification up to 300~V. Liquid nitrogen-cooled Si and
Ge detectors were used for the detection of the second harmonic ($2f$)
component. No signal was measured at $f$ and $3f$. We experimentally obtained
the variation of transmission $- \Delta T$ which could be related to the change
in absorption $\Delta \alpha$ by the relation: $\Delta \alpha \sim \frac{-\Delta
T}{T}$.

\section{Results and Discussion}
\label{results}
\begin{figure}
    \centering
    \resizebox{\columnwidth}{!}{
	\includegraphics[width=8.6cm]{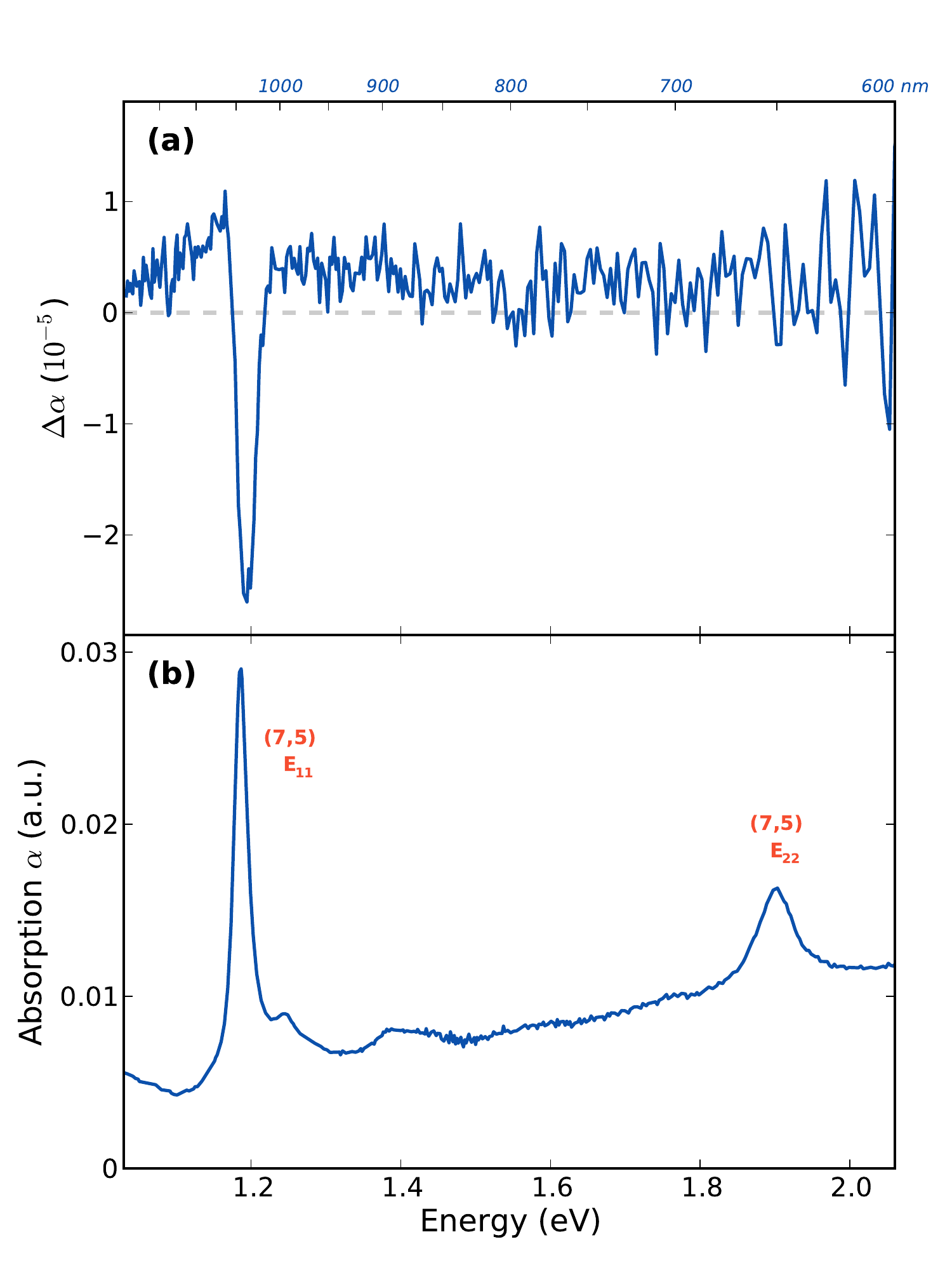}
    }
	\caption{(color online) Electroabsorption variation (a) and optical absorption (b)
	spectra of CoMoCAT/PFO extracted sample. Absorption peaks at 1.18~eV and
	1.9~eV correspond respectively to $E_{11}$ and $E_{22}$ excitonic
	transitions of (7,5) SWNT. The electroabsorption spectrum was performed
	at a field of 12~kV/cm.}
	\label{Fig1}
\end{figure}

Room temperature electroabsorption spectrum (EA) of CoMoCAT/PFO at 12~kV/cm is
reported in Fig.~\ref{Fig1}a, while the corresponding absorption spectrum is
shown in Fig.~\ref{Fig1}b. A striking feature in the EA spectrum is the
unique and well defined peak at 1.18~eV. Interestingly, there is a corresponding
feature in the absorption spectrum, also at 1.18~eV. The CoMoCAT/PFO absorption
spectra is mainly constituted by two main peaks at 1.18~eV and 1.9~eV. Those
peaks correspond to the $E_{11}$ and $E_{22}$ optical exciton transitions below
the $\Delta_{11}$ and $\Delta_{22}$ continuums. Using well-known assignation
rules\cite{Science-Oconnell}, one can assign the observed excitonic
transitions to index-defined nanotubes. In the case of CoMoCAT/PFO sample, peaks
at 1.18~eV and 1.9~eV could respectively be attributed to the (7,5)'s $E_{11}$
and the (7,5)'s $E_{22}$ excitons. The other two small features around 1.24~eV
and 1.4~eV are attributed to trace of different nanotubes, probably (6,5) and
(6,4) s-SWNT. Therefore, the observed peak in the EA spectrum could be assigned to the
(7,5) SWNT's first exciton.

\begin{figure}
    \centering
    \resizebox{\columnwidth}{!}{
	\includegraphics[width=8.6cm]{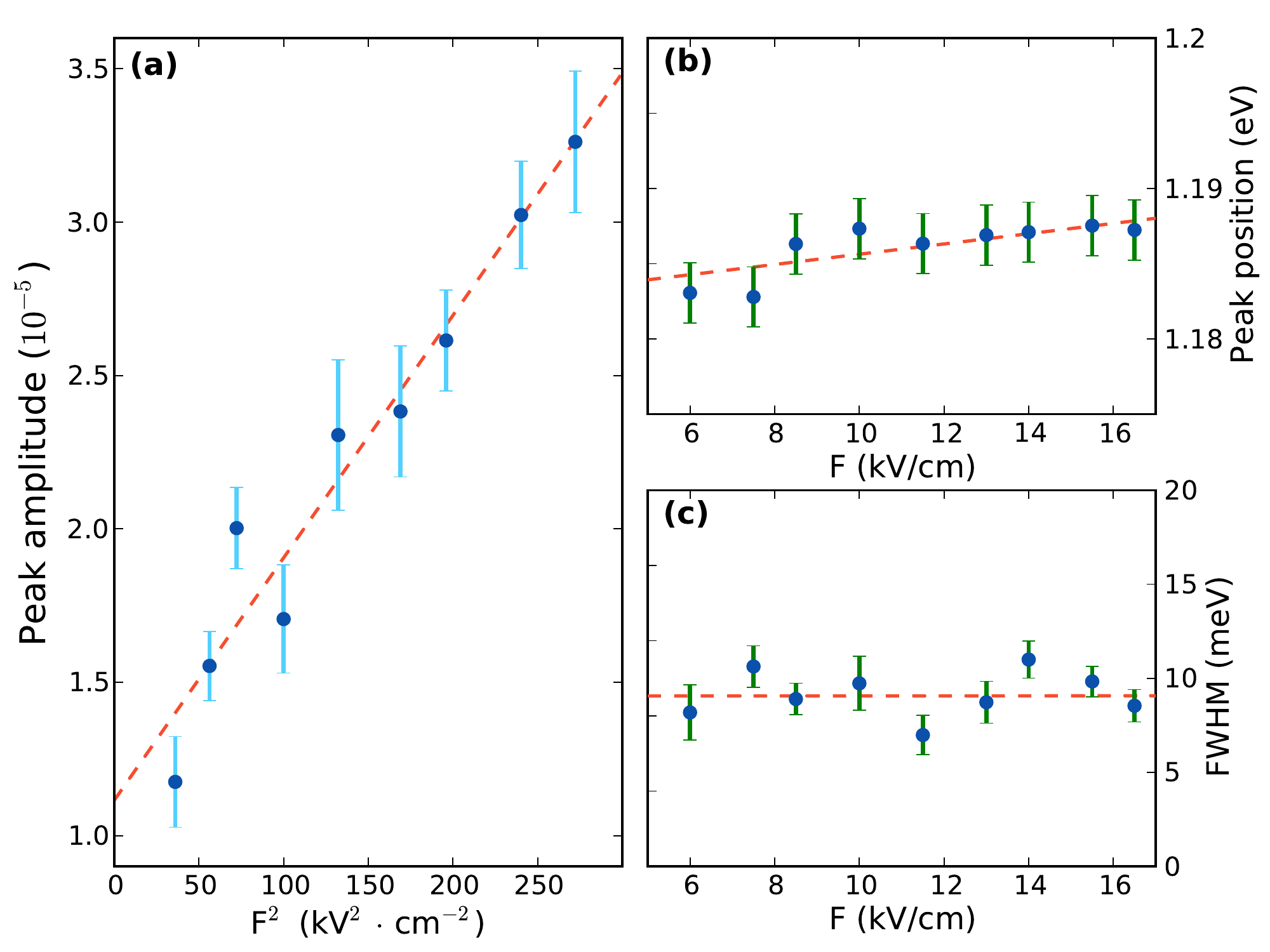}
    }
	\caption{(color online) Amplitude $\Delta \alpha$ (a), position (b), and full width half
	maximum (FWHM) (c) as a function of the electric field applied of (7,5)
	SWNT electroabsorption peak in CoMoCAT/PFO extracted sample. Errors bars
	are the standard 2~$\sigma$ deviation from the lorentzian fit, while the
	dashed line is the linear regression.}
	\label{Fig2}
\end{figure}

In order to further investigate the origin of this EA peak, a series of EA
spectrum were performed on sample CoMoCAT/PFO, while varying the applied field
from 6 to 16.5~kV/cm. Below 6~kV/cm, the EA signal was hardly distinguishable
from noise, while above 16.5~kV/cm the voltage amplification system saturates.
The (7,5) s-SWNT EA peak was fitted by a single lorenzian, and amplitude $\Delta
\alpha$, peak position, and its full width half maximum (FWHM) were extracted.
Peak amplitude as a function of the square electrical field $F^2$ is shown on
Fig.~\ref{Fig2}a, while peak position and FWHM as a function of the electrical
field $F$ are shown in Fig.~\ref{Fig2}b and c, respectively.

No significant peak shift or peak broadening could be detected while applying
the electrical field from 6 to 16.5~kV/cm. This means that the field applied to
the system remains low, and that the excitonic dissociation is negligible. In
that case, EA theory\cite{prl-Zhao,nano-Perebeinos} states that
$\Delta \alpha$ obey the following relation:
\begin{equation}
    \Delta \alpha = K_1 \cdot \frac{d_t^2}{E_b^2} \cdot F^2 + K_2 \label{eq1}
\end{equation}
where $E_b$ is the excitonic binding energy, $d_t$ the SWNT diameter, $F$ the
applied electric field, and $K_1$ and $K_2$, constants. As in the current case,
there is only one kind of nanotube, $E_b$ and $d_t$ are both constant, and
$\Delta \alpha$ should be linear with $F^2$, which is what it was experimentally
obtained (Fig.~\ref{Fig2}a). This demonstrate that the observed EA peak at
1.18~eV in Fig.~\ref{Fig1}a is from excitonic origin, as it was also observed by
Kishida et al.\cite{prl-Kishida}

\begin{figure}
    \centering
    \resizebox{\columnwidth}{!}{
	\includegraphics[width=8.6cm]{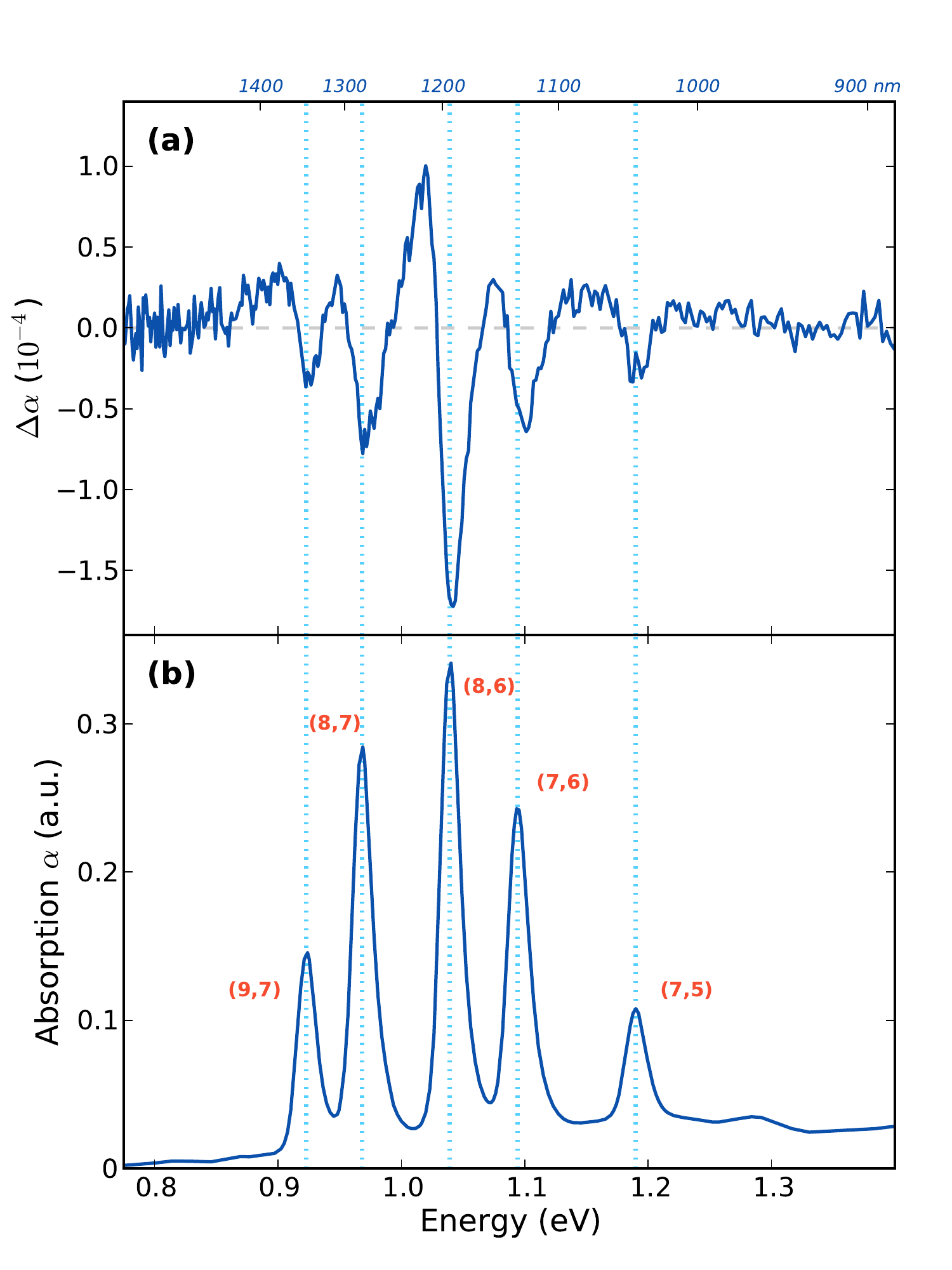}
    }
	\caption{(color online) Electroabsorption (a) and optical absorption (b)
	spectra of HiPCO/PFO extracted sample. Absorption peaks at 0.923, 0.968,
	1.039, 1.094 and 1.19~eV correspond respectively to (9,7), (8,7), (8,6),
	(7,6) and (7,5) SWNT $E_{11}$ excitonic transitions. The
	electroabsorption spectrum was performed at a field of 12~kV/cm.}
	\label{Fig3}
\end{figure}

Relations between nanotube diameter and EA spectra were also investigated.
Previously prepared HiPCO/PFO sample, which contains several well identified
s-SWNT chiralities with well separated absorption features, is ideal for that
study. Room temperature EA spectrum at 12~kV/cm and the corresponding absorption
spectrum are reported in Fig.~\ref{Fig3}.

EA spectrum is mainly composed of five peaks, with five corresponding peaks in
the absorption spectrum. As it was previously demonstrated, it is possible to
assign optical transitions to individual s-SWNT $E_{11}$ excitons, ranging from
(7,5) to (9,7) s-SWNT, with diameter ranging from 0.83 to 1.10~nm. As the
absorption peaks of the HiPCO/PFO sample are well separated from each other, the
resulting EA spectrum is constituted by the individual contribution from those
five s-SWNT, making the assignment of EA peaks straightforward. An asymmetry
could be observed on all EA peaks in the HiPCO/PFO sample, as well as to a
lesser extent in the CoMoCAT/PFO sample. This effect is particularly clear with
the (8,6) EA peak at 1.039~eV, which is very intense with a clear asymetric
feature at lower energy. This asymetric feature in EA peaks is expected by
theoretical results from Zhao et~al.\cite{prl-Zhao}, and our experimental
results are in good agreement with those calculations.

\begin{figure}
    \centering
    \resizebox{\columnwidth}{!}{
	\includegraphics[width=8.6cm]{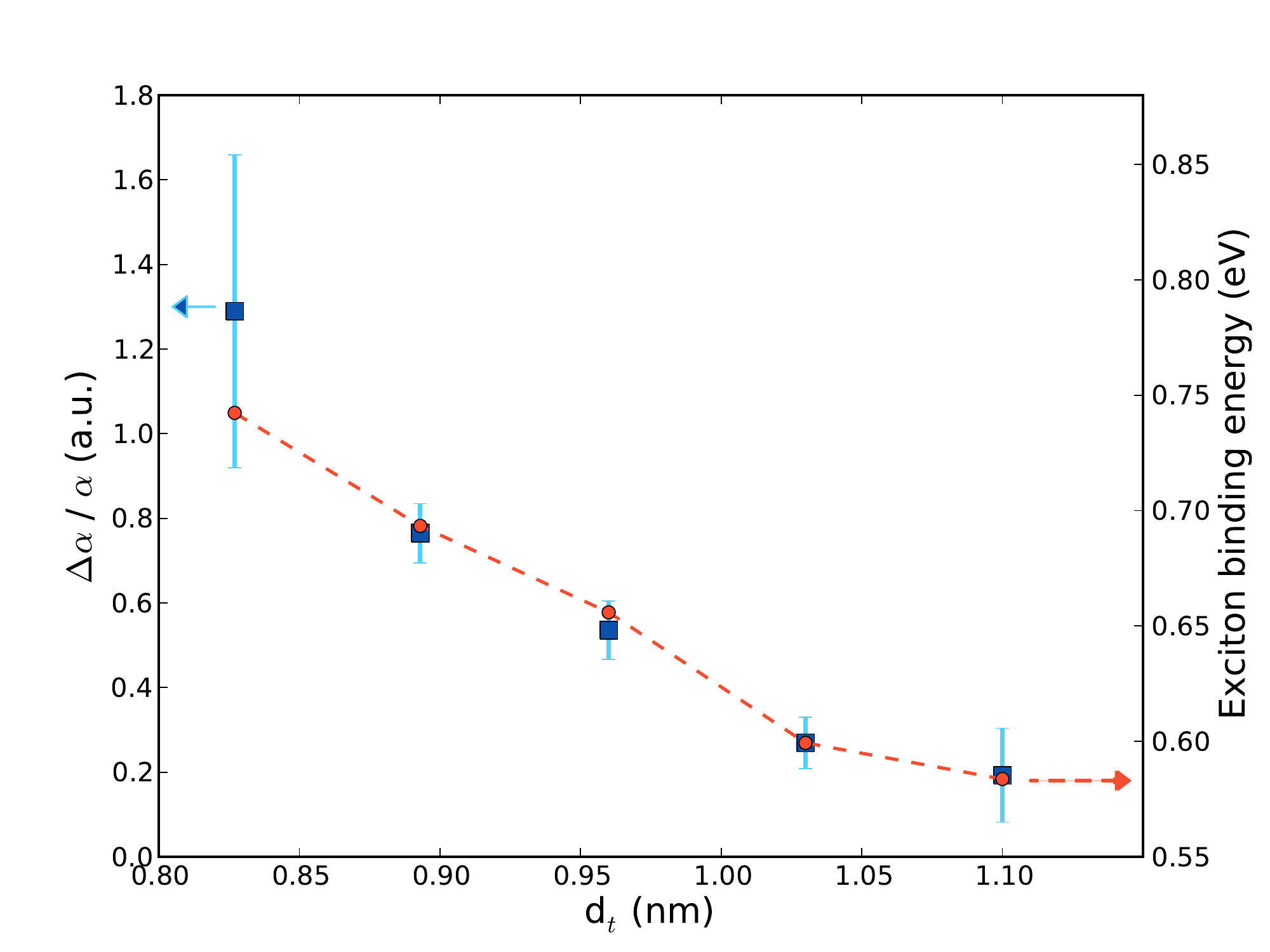}
    }
	\caption{(color online) Normalized electroabsorption response $\Delta
	\alpha / \alpha$ of HiPCO/PFO extracted nanotubes (square, left) and excitonic
	binding energy $E_{b}^{11}$ (circle, right) as a function of the nanotube
	diameter $d_t$.}
	\label{Fig4}
\end{figure}

An interesting way to analyze EA relations with s-SWNT diameter consists on
normalizing the EA amplitude $\Delta \alpha$ at a given field by the absorption
amplitude $\alpha$ for each nanotube. Indeed, this gives the intrinsic EA
response of each s-SWNT independently from its concentration in the samples.
Plot of this normalized EA amplitude as a function of the nanotube diameter is
reported in Fig.~\ref{Fig4}.
Exciton binding energy $E_b$ could be given by the following analytical
expression from Capaz et~al.\cite{prb-Capaz}:
\begin{equation}
	E_b = \frac{1}{d_t} \left( A + \frac{B}{d_t} + C \zeta + D \zeta^2
	\right)
	\label{eq2}
\end{equation}
where $d_t$ is the tube diameter, and $\zeta = (-1)^\nu \cos 3\theta / d_t$
represents the chirality dependence. $A$, $B$, $C$ and $D$ are numerical
constants, and their values are respectively\cite{prb-Capaz}:
\begin{eqnarray*} 
  A &= &0.6724 \textrm{ eV nm} \\
  B &= &-4.910 \cdot 10^{-2} \textrm{ eV nm}^2 \\
  C &= &4.577 \cdot 10^{-2} \textrm{ eV nm}^{-2} \\
  D &= &-8.325 \cdot 10^{-3} \textrm{ eV nm}^3
\end{eqnarray*}

Exciton binding energy for the five s-SWNT in which we had measured normalized
EA amplitude were calculated using Eq.~\ref{eq2} and reported in
Fig.~\ref{Fig4}. Normalized EA response $\Delta \alpha / \alpha$ is in very good
agreement with the calculated exciton binding energy $E_{b}^{11}$. However, this
relationship currently holds true only for a limited extent of nanotube
diameters and exciton binding energies. In particular, s-SWNT with diameter
above 1.10~nm and exciton binding energy below 550~meV may not follow this rule.
Even if challenging, additional experiments needs to be performed on higher
diameter s-SWNT to extend the boundaries of the current study. This could allow
establishing a definite connection between electroabsorption and exciton binding
energy. Nevertheless, electroabsorption spectroscopy is a powerful technique to
directly probe carbon nanotubes excitonic states, and it may become a useful
experimental tool to investigate the modification of s-SWNT's excitons under
changes in surrounding medium or chemical functionalization.

\section{Summary}
\label{sum}
We report the first electroabsorption experiments performed on well identified
index-defined semiconducting nanotubes. The electroabsorption response is
sufficiently defined to allow a peak-to-peak correspondence with the absorption
spectrum, leading to an unprecedented indexation of electroabsorption spectrum.
Due to its excitonic nature, electroabsorption amplitude is quadratic with the
applied electric field. Our results show that at least for a limited range of
diameter, electroabsorption is proportional to exciton binding energy, in good
agreement with theoretical works. Electroabsorption is an interesting technique
to directly probe excitons in carbon nanotubes. From that aspect,
electroabsorption proves to be an invaluable tool for the direct study of
excitonic-based effects in future carbon nanotubes-based photonics devices, and
in particular, to develop optical modulators based on electroabsorption effects.
\begin{acknowledgement}
N. Izard thanks the Japan Society for the Promotion of
Science and CNRS for financial support.
\end{acknowledgement}

\bibliographystyle{epj.bst}
\bibliography{electroabs.bib}

\end{document}